\documentclass[prl,twocolumn,showpacs,amssymb,superscriptaddress]{revtex4}

\usepackage{graphicx,dcolumn,amsmath}

\begin{document}

\title
{ 
Superconductivity in
carbon nanotubes coupled to
transition metal atoms.
}
\author
 {Nacir Tit}\email[National Research Council visiting scientist: 
Email address\ ]{ntit@uaeu.ac.ae}
\affiliation{
Dept. of Physics,
UAE University, P. O. Box 17551, Al-Ain, United Arab Emirates}
\author{\firstname{M. W. C.}
\surname{Dharma-wardana}} \email[Author to whom 
correspondence should be
addressed: Email address:\ ]
{chandre@nrcphy1.phy.nrc.ca}
\affiliation{
National Research Council, Ottawa, Canada. K1A 0R6\\
}

\date{\today}
\begin{abstract}
The electronic structures of zig-zag and arm-chair single-walled
carbon nanotubes interacting with a 
 transitional-metal atomic nanowire  of Ni
 have been determined.
The Ni nanowire
creates a large electron density of states (DOS)
at the Fermi energy. The dependence of the enhanced DOS
on the spin state and positioning of the transition-metal wire
 (inside  or
outside the nanotube) is studied.
 Preliminary estimates of the electron-phonon
interaction
 suggest that
such systems may have  a superconducting transition
temperature of $\sim$ 10-50 K.
The signs of superconductivity seen in ``ropes'' of
nanotubes may also be related to  the effect of intrinsic
transition-metal impurities.
\end{abstract}
%
\maketitle
%
 The discovery of carbon nanotubes (CNT) has given rise to  fascinating
 basic physics  as well as
tantalizing  technological
possibilities\cite{mit}.
The synthesis of CNTs naturally incorporates transition-metal
(TM) atoms like Ni, Co, which are used as 
``seeds'' for initiating growth. 
In addition, TM impurities are found in CNTs, as
adsorbed species  inside or outside the CNT walls. 
TM atoms occurring as substitutional impurities
are energetically unfavourable, and annealing 
converts them into adsorbed species\cite{andriotisMD}.
Many  studies of CNTs with alkali, Au, Ti, Co, Ni, etc., and rare earth
atoms,
have appeared in the literature and have clarified the binding energies
and other aspects of these systems\cite{mit,meunier}.
%
Given the strong $d-d$ interaction of TM atoms, and the directionality
imposed by the CNTs, the TM atoms
 may form quasi one-dimensional 
 transition-metal nano-wires (TMWs), 
 on the inner or outer  surface
of the CNTs\cite{yang}. 
 From a theoretical point of view, the 1-D structure of the CNT, 
and the 1-D TMW, provide a realization of
Luttinger liquids, Bethe ansatz problems, and novel spintronics.


The TMW transfers electrons to the CNTs and can transform semi-conducting
CNTs into metals. If the electron density of states (DOS) at the Fermi
surface could be strongly enhanced,
the coupled system may respond by undergoing a Peierls 
or a  Cooper-pairing transition.
In fact, the possibility of superconductive interactions in pure CNTs
themselves has been discussed, mainly within 
the Luttinger-liquid paradigm\cite{gonzales}.
Evidence for superconducting fluctuations, proximity effects etc.,
 in CNTs have been reported
\cite{tang}.
 The quasi 
1-D nature of the TM chain, and the CNT are modified by 
coupling them  when electrons can hop from the CNT to the TMW and back.
The Cooper pairing would involve electronic and
phononic effects associated with the Ni-C as well as the Ni-Ni and C-C
interactions. Even if these  do  not lead to superconductivity,
we may expect  some interesting and technologically exploitable
properties from the TMW/CNT system.

	The objective of this paper is to use tight-binding (TB) schemes,
supported by density-functional  first-principles calculations,
for calculating the electronic energy bands,
 the density of states, and phonon properties
of CNTs coupled to TMWs.
The TB results are 
used for a preliminary discussion of the possibility of
superconductivity in these systems.
 It is concluded that coupled
CNT/TMW systems 
are promising superconductive materials whose special
properties may be valuable in a variety of novel applications.

We use TB parameters 
 obtained from fitting to first-principles calculations,
or by adapting from  Harrison's universal parameters\cite{harrison}
 when  justifiable.
Our simulation cells (SC) contain one or two TM atoms and
also one or two CNT unit cells, as the case may be.
 Single-walled CNTs are specified by the pair
of numbers (n,m) which defines the chirality of the nanotube. 
These numbers specify the way  a 2-dimensional graphite sheet 
is rolled to obtain the CNT. While the $\pi$- electron
system in the generic graphite sheet makes it conducting, the metallic 
or semiconducting nature of the CNT (i.e, its bandgap) is
determined by the chosen (n,m) configuration.
In this study we consider the
two extreme sets, (n,0) and (n,n), which correspond to the 
 zig-zag (ZZ) and arm-chair (AC) configurations.

\begin{figure}
\includegraphics*[width=9.0cm, height=9.0cm]{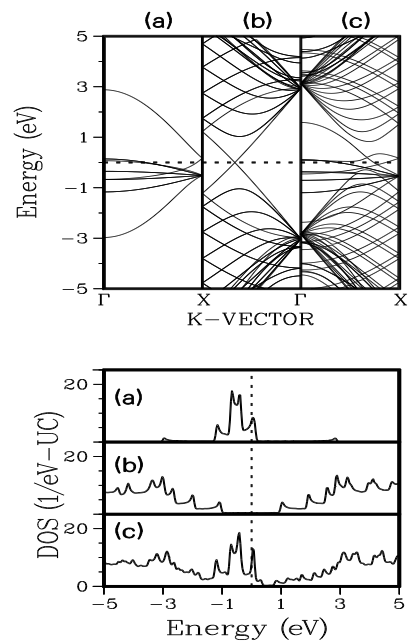}
\caption
{ Energybands  and DOS of (a) chain of 
ferro-magentic Ni atoms, (b) arm-chair CNT (9,9)
 and (c) the coupled system Ni-wire/CNT
with the TMW adsorbed on the outside wall.
The DOS is per eV per simulation cell (SC)
with 72 carbon atoms and 2 Ni atoms.
}
\label{cntarm}
\end{figure}

\begin{figure}
\includegraphics*[width=9.0cm,height=9.0cm]{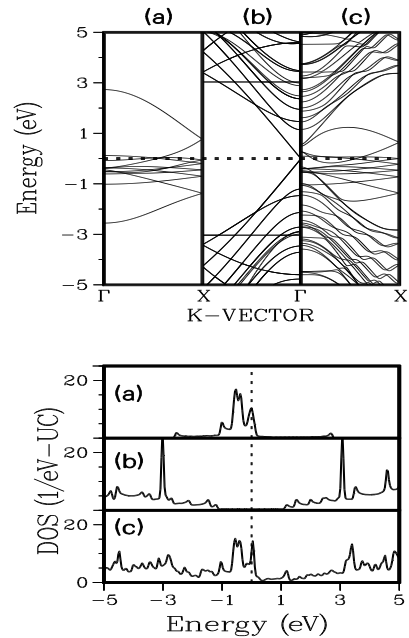}
\caption
{ Energybands  and DOS of (a) chain of
ferro-magentic Ni atoms, (b) zig-zag CNT (12,0)
 and (c) the coupled system Ni-wire/CNT with the
TMW adsorbed on the outside wall.
The DOS is per eV per simulation cell (SC)
with 48 Carbon atoms and 2 Ni atoms. 
}
\label{cntzig}
\end{figure}

 The use of two Ni atoms per simulation cell allow us to consider
  ferro- or antiferromagnetic nearest-neighbour
coupling. If one TM atom per SC were used,
 the Ni-Ni interaction is negligible and
we have the case of doping with almost isolated TM atoms.
	A first-principles study of a 1-D chain of Ni atoms in isolation
has been carried out by Freeman and colleagues\cite{freeman},
 and
used to obtain the tight-binding parameters for the Slater-Koster
method. While the diagonal parameters are strongly
modified from the ``universal'' parameters of Harrison, the off-diagonal
universal parameters  are quite close to the results obtained from the
fit to the first-principles calculations. The TB parameters are
further evolved into a set of parameters which depends on the antiferromagnetic
or ferromagnetic coupling between Ni neighbours
by making the on-site energy of the
Nickel $d$-orbital a function of the spin state of the Ni atom. Thus we
have used  $\epsilon_d\uparrow-\epsilon_d\downarrow$ = $J$ with a value
 of $J$ corresponding to the bulk Ni value. A better estimate, suitable for
the 1-D chain may be obtained by fitting to a first-principles calculation,
 but the results are found to be rather insensitive to the
spin configuration and may be considered
illustrative. The bandstructure and DOS of the 
Ni wire (uncoupled from the CNT) are seen in panel (a)
 of Fig.~\ref{cntarm}. Here the, Ni wire which registers with the
armchair CNT is a linear chain. The Ni wire which registers with the
zigzag CNT is itself a zigzag chain. Its bandstructure and DOS before
coupling to the CNT are shown in panels (a) of
Fig.~\ref{cntzig}.

In panels (b) of Figs.~\ref{cntarm},~\ref{cntzig} we show the bandstructure
and DOS of the CNTs before coupling to the TMWs.
 The Slater-Koster approach for obtaining the electronic
states of CNTs within a $sp^3$ basis is used.
Many TB calculations for CNTs
have been reported and the procedure  is now
well established.
The AC(9,9) CNT, panel (b) of Fig.~\ref{cntarm},
 has two bands crossing the Fermi energy
near 2/3 of the  $\Gamma-X$ line and hence
it is metallic. The ZZ systems become metallic for
specific values of n (for multiples of 3), and the bandgap
for the semiconducting case is,
to first order, a function of $1/d$. Here $d$ is the ideal
CNT diameter given by $d=(a_L/\pi)(m^2+m^2+mn)^{1/2}$, where 
$a_L=0.245$ nm is the lattice constant in the 2-D graphene sheet.
Panel (b) shows a very small flat DOS at $E_F$ for the isolated
CNT. Figure~\ref{cntzig} shows, in panels (b) the bandstructure
and DOS of the zig-zag (12,0) nanotube without coupling to the
TMW. Here also, although the
nanotube is metallic, its small DOS at $E_F$ is hardly visible in
panel (b).

We position the nanowire inside, or outside
the CNT, and interacting with the carbon $s$ and $p$ 
bonded network forming the
CNT wall. 
The TMW binds chemically to
the CNT, unlike in the case of, e.g., Au or Al. The interaction of Ni atoms
with CNTs have also been studied by Andriotis et al.\cite{andriotisMD},
who
determined the relaxation of substitutional and 
also adsorbed Ni atoms near (or inside)a single-walled
CNT. Their simulations show that a Ni atom may migrate into
a CNT, from the outside, using a vacancy in the CNT wall.
These calculations, as well as other studies,
 already provide us with information regarding the
optimal Ni-C distance ($\sim$ 0.2 nm). For the two-Ni
SC we have taken the Ni-Ni bond
to be nearly the same as that of the isolated nanowire\cite{freeman}, and 
registering with the $a_L$ distance on the CNT wall.
 This is consistent with the
fact that the local structure of the CNT is only minimally affected by the
presence of an adsorbed Ni atom inside or outside the CNT wall. However, as
seen from our calculations, a large electron density of states is
created at the Fermi point by the Ni nanowire, and the CNT becomes
conducting, even if it were originally semiconducting (however,
this is not always the case in Ti nanowires). The results for the
coupled CNT/TMW are given in panels (c) of Figs.1-2.

	The bandstructure of the Ni-wire/CNT system, panel (c), shows
a nearly dispersionless (``flat'') band  which crosses 
the Fermi energy, and also other Ni-like
bands which show greater dispersion. The ``flat'' band relates to
 electron hopping via the Ni-C bond which is 
localized. It provides transport via hopping from the CNT
 to the TMW and back.  The bands with linear dispersion
 found at the Fermi energy of 
isolated AC nanotubes (used in bosonization approaches) are no longer
present in the coupled CNT/TMW systems.
Note that the linear-k bands
crossing $E_F$ in panel (b) have
been repelled apart in the coupled AC system (c) 
and the bands derived from the Ni atoms are inserted
near the Fermi energy. Also, the doubly degenerate bands in panel (b) are
 split in panel (c).

The density of states of the Ni chain, the (9,9)CNT and the coupled
system are shown in Fig.~\ref{cntarm}. The main change in the DOS is the 
enhancement at and around the Fermi energy. The splitting of the 
bands has smoothened  the DOS, as seen by comparing panels (b) and (c).
Similar results are found in the density-functional calculations
of Yang et al for  metallic Ti nanowires coupled to CNTs.

%
Our calculations, and those  of Yang\cite{yang} for Ti 
 show that there is significant
electron transfer between the CNT and the TM. The density of states
$N(\epsilon_F)$ for one-Ni atom in the simulation cell ( i.e, isolated
Ni atoms) is  three to four times stronger than the case with two-Ni
atoms per cell. The TM wire retains some of its electrons on the wire,
while the isolated Ni atoms inject more electrons into the CNT
(this needs further confirmation from a more microscopic calculation).
On the other hand, if one Ni atom were used in  a simulation cell of 3 units,
then the doping level would be lower and the effect is less.
This implies that about 3-4\% of Ni atoms in (9,9) or (12,0) nanotubes
would be optimal for a highly enhanced $N(\epsilon_F)$.


	The presence of a significant density of states at the Fermi
energy opens the possibility of structural relaxation via the electron-lattice
interaction. This was in fact considered by Mintmire et al., and 
also Saito et al.\cite{mintmire},
%
%
who concluded that the CNTs are
stable with respect to Peierls-type distortions.
%
%
The
density-functional total-energy minimisations of
Yang et al.\cite{yang} for CNTs coupled to Ti nanowires
show only  a slight local modification of the CNT lattice.
 Our calculations agree
with this and the main effect arises from electron transfer and
self-consistent  readjustment
of the Fermi energy. Hence we conclude
that, even if isolated CNTs are unstable,
the coupled CNT/TMW systems are stable with respect to Peierls distortions,
and that the enhanced DOS, $N(\epsilon_{F})$ at the Fermi energy 
is available for modification by more subtle
processes like Cooper pairing.

 Graphite intercalation compounds (GIC) have 
 very low  transition temperatures $T_c\sim 1$ K, and $N(\epsilon_F)$
values of $\sim$ 1-2 states/eV-spin for 60 Carbon atoms, and compares
with the $N(\epsilon_F)\sim 10$ states/eV-spin-C$_{60}$ of alkali-doped
fullerenes (ADF).  Superconductivity in the ADFs
have been discussed using purely electronic mechanisms\cite{kievelson},
but mostly using the  McMillan-Eliashberg (ME) approach within the
Migdal approximation\cite{schluter}.
Migdal's approximation is more easily justified in the CNT/TMW systems
for processes mediated by the heavy transition-metal atoms.
Schluter et al.\cite{schluter}, argued
 that the electron-phonon interaction in the ADFs is larger than in
GICs because of the curvature effect absent in the GICs\cite{schluter}. 
This argument applies equally to CNTs.
The superconductivity in ADFs involves not only the ``on-ball'' processes,
but also the much weaker hopping in a 3-D system of fullerene molecules.
In  studies on superconductivity of CNTs, the quasi 1-D character is
claimed to be  overcome by weak hopping between single
CNTs in ``ropes'' of nanotubes. This hoping involves tunnelling through
non-bonding interactions and is very poor.
In fact, calculations by Delaney et al.\cite{delaney}
show that the
non-bonding interactions suppress the 1-D conductivity.
On the other hand, in CNT/TMW coupled systems, electron transfer 
between two nanotubes  occurs effectively  via physical
metallic contact between  TMWs on adjacent CNT/TMW strands. Hence the
systems aquire 3-D character and the 
main issue is to study the nature of the interactions on the individual
CNT/TMW elements. 


A force-constants model for the CNTs\cite{saitoP},
extended to include  Ni-Ni and Ni-C interactions can be used to
study the phonons of the CNT/TMW system. 
Depending on the force constants we use, the high frequency Ni-Ni vibrations
fall between 200-300 cm$^{-1}$ and  strongly interact with the 
``breathing'' and ``twisting'' modes of the CNTs, while leaving the
carbon high-energy optical modes relatively unaffected, 
except for lifting all degeneracies. 
The axial
symmetry of the isolated CNT is broken in the coupled system, and a rich
redistribution of the low-energy phonon modes is found.\cite{phonons}
The radial modes are not too important in ADFs since they have
little effect on the electronic structure at the Fermi level. 
In contrast, radial modes in the CNT/TMW systems couple strongly with the 
Fermi level bandstructure, and are  many times more efficient
in  electron-phonon (e-p) coupling
than the stiffer tangential modes. 
The e-p interaction associated
with electron hopping between the CNT and the TMW, and along the
TMW would bring new, 
strong coupling features absent in the pure CNTs. Electron transport in the
Ni bands would also be subject to strong e-p interactions.
The electron-phonon coupling constant is  of the form
\begin{equation}
\label{tc}
\lambda=N(\epsilon_F)V=N(\epsilon_F)\sum_i \eta_i/<M_i\omega_i>
\end{equation}
 where the sum runs over
all vibrational modes, with atomic masses $M_i$, and $\omega_i$ and $\eta_i$
being averaged contributions from phonon frequencies and
Hopfield factors for the  electronic states.
The $N(E_F)$ depends somewhat on the Ni-spin configuration (ferro or
antiferro), the location of the TMW (inside or outside the CNT), and the
nature of the CNT (ZZ, AC), as seen in Table I, but a ``grosso modo'' value is 
0.035 states/(eV-spinstate).  To clarify the notation we have used here,
(12,0)F$Ni_2$ indicates a calculation for a ZZ nanotube with two Ni atoms
in a simulation cell containing 1 unit cell of the ZZ-CNT. Hence there
 are 48 carbon atoms, with 192 states, and each Ni atom has one $s$ and
5 $d$ states. The total number of spin-states is 408 per simulation
cell, yielding 
a total $N(\epsilon_F)$ of about 12 states per eV per simulation cell.
The number reported in the Table is the $N(\epsilon_F)$ per eV per spin-state.
In the (9,9)AC system we use one unit cell in the simulation cell,
giving 72 carbon atoms and two Ni atoms.
The normalization 
 enables us to compare the different structures given in Table I.
 When there is only one Ni atom per
SC, there are no Ni-Ni interactions and the Ni electrons
are (3-4\% doping) most effective 
in augmenting the $N(\epsilon_F)$ in ZZ nanotubes.
\begin{table}
\caption{electron density of states per eV per spin per electron
in CNT/TM coupled systems. With two Ni atoms in the simulation cell, 
Ferromagnetic (F) and  antiferromagnetic (AF) alignments are given.
With one Ni atom/simulation cell, there are no Ni-Ni interactions.
}
\begin{ruledtabular}
\begin{tabular}{lcccccccc}
system&(12,0)&(13,0)&(9,9)\\
\hline\\
F $\;\;Ni_2$ in & 0.03 & 0.035 & 0.015\\
AF$\;Ni_2$ in & 0.04 & 0.03  & 0.02  \\
F $\;\;Ni_2$ out&0.035 & 0.03  & 0.015 \\
AF$\;Ni_2$ out&0.036 & 0.033 & 0.019 \\
$\;\;\;Ni_1$ out  &0.08   & 0.08  &0.06   \\
\end{tabular}
\end{ruledtabular}
\label{table1}
\end{table}

The 0.035 per eV per spin-state as defined previously
 is comparable to that in ADFs where $N(E_F)$ of 
 2-8 states/(eV-spin-$C_{60}$) have been quoted\cite{kumar,schluter}
%
The $T_c$ depends on the Coulomb parameter $\mu*$. Koch et al.\cite{koch}
 have
argued that $\mu*$ is $\sim$ 0.1-0.2 even though the reduction due to
retardation effects is absent in fullerene-like systems. Hence, using this
range of parameters, the $T_c$ of CNT/Ni-wire coupled systems is
found to be in the range 10-50 Kelvin.

	A natural implication of our study is that the observed signs of
 CNT superconductivity\cite{tang}, e.g., in ``ropes'' of
CNTs may be due to intrinsic doping by TM impurities which
cannot be removed completely\cite{dujardin}.
 This can be examined
via experiments in which the TM impurities are added or 
leached out in a controlled way.
When isolated Ni atoms are attached, e.g, one Ni per 48 carbon
 atoms in a (12,0) CNT, the
$N(\epsilon_F)$ is enhanced. The random Ni-Ni contacts between
such CNTs in ropes of CNTs
could provide the higher dimensionality needed to stabilize the
superconductivity. In conclusion, our calculations on CNT/TM systems suggest
that couplings to suitable TM  atoms
 stabilize any inherent superconductivity in the CNTs. These
conclusions, based on tight-binding methods should stimulate more
microscopic calculations as well as new experiments.

%

%
%

%


\begin{references}
\bibitem{phonons}
A detailed discussion of the phonons in the CNT/TMW system will be
presented elsewhere.
\end{references}
\end{document}